\documentstyle[psfig2]{jpp}

\def\ee{\end{equation}}
\def\be{\begin{equation}}
\def\bdm{\begin{displaymath}}
\def\edm{\end{displaymath}}

\def\sa{{\rm sgn}(q_a)}
\begin{document}
\title[Kinetic theory of interstellar density and magnetic field fluctuations]
{Relations between interstellar density and magnetic field fluctuations I. 
Kinetic theory of fluctuations}
\author[R. Schlickeiser and I. Lerche]{R. \ns  
S\ls C\ls H\ls L\ls I\ls C\ls K\ls E\ls I\ls S\ls E\ls R
and 
I.\ns L\ls E\ls \ls R\ls C\ls H\ls E\footnote{Permanent address:
Department of Geological Sciences, University of South Carolina, Columbia, SC 29208, USA; black@geol.sc.edu}}
\affiliation{Institut f\"ur Theoretische Physik, Lehrstuhl IV: 
Weltraum- und Astrophysik, Ruhr-Universit\"at Bochum, 
D-44780 Bochum, Germany}
\date{18 March 2002, in revised form 29 May 2002}
\maketitle
\begin{abstract}
Using linear kinetic plasma theory the relation between 
electron density and magnetic field
fluctuations for low-frequency plasma waves for Maxwellian background distribution functions
of arbitrary temperatures in an uniform magnetic field is derived. 
By taking the non-relativistic temperature limit 
this ratio is calculated for the diffuse intercloud medium in our Galaxy.
The diffuse intercloud medium
is the dominant phase of the interstellar medium with respect to
radio wave propagation, dispersion and rotation measure studies. 
The differences between the relation of electron density and magnetic field
fluctuations from the linear kinetic theory as compared to the classical MHD theory
are established and discussed.
\end{abstract}

\section{Introduction}
Important input quantities for the quasilinear test-particle description of 
cosmic ray transport in weakly-turbulent astrophysical plasmas are the wavenumber
power spectra of magnetic field fluctuations. Within the plasma wave viewpoint, the plasma
irregularities are usually modelled as a superposition of linear waves well below the ion 
cyclotron frequency, such as Alfven and magnetosonic waves. However, the observed 
turbulence properties in the more distant interstellar and intergalactic plasmas are obtained
from radio propagation measurements as dispersion measures, rotation measures and
interstellar scintillation. These turbulence diagnostics are biased towards the high-density 
ionized interstellar phases
with large volume filling factors, i.e. the diffuse intercloud gas and HII envelopes. In
particular, dispersion measure and scintillation data are primarily diagnostics of density
and only secondarily of magnetic field. These diagnostics demonstrate the existence of
interstellar density irregularities with Kolmogorov-type ($ \propto \omega ^{-s}$, $s=const.$)
frequency power spectra extending over 11 decades 
in frequency, much below ($\omega <<\Omega _p$) the proton gyrofrequency (Rickett \cite{r90}, 
Armstrong et al. \cite{ars95}). 

Often the electromagnetic fluctuations are described within magnetohydodynamic (MHD) theory
(e.g. Sturrock \cite{s94}, Goldreich \& Sridhar \cite{gs95}, 
Hollweg \cite{h99}, Lithwick \& Goldreich \cite{lg02}) 
which is appropriate at large turbulence scalelength $l\ge l_{MHD}$.
However, the plasma parameter of the diffuse intercloud medium 
$g=\nu _{ee}/\omega _{p,e}\simeq 10^{-10}$ is much smaller than unity, so that a kinetic description of
the electromagnetic turbulence seems to be necessary. It is the purpose of
the present paper to provide the relation between electron density and magnetic field
fluctuations on the basis of the linear kinetic plasma theory. 
In paper II of this series we
will use the kinetic turbulence relations to calculate frequency power spectra of electron density 
fluctuations 
from anisotropic power spectra of magnetic field fluctuations in the form of Alfven and
magnetosonic waves. Such anisotropic interstellar magnetic field power spectra are required in
order to be in accord with the heating/cooling balance of the diffuse intercloud medium 
(Lerche \& Schlickeiser \cite{ls1}).
\section{Diffuse intercloud plasma parameters}
According to 
Rickett (\cite{r90}), Spangler (\cite{s91}), Lerche \& Schlickeiser
(\cite{ls1}) the diffuse intercloud gas can be well modelled 
as an isotropic Maxwellian plasma with the following plasma and turbulence parameters: 
$B_0=1.3 \mu $G, 
$n_e=0.08$ cm$^{-3}$, $s=5/3$, $v_e\simeq 2\cdot 10^7$ cm s$^{-1}$, $T_e=T_p$,
$v_p=v_e (m_e/m_p)^{1/2}=4.7\cdot 10^5$ cm s$^{-1}$, 
$k_{min}=2\pi/l_{max}=2\cdot 10^{-18}$ cm $^{-1}$ 
with $l_{max}=3\cdot 10^{18}$ 
cm, $k_{max}=2\pi/l_{min}=8\cdot 10^{-8}$ cm$^{-1}$ 
with $l_{min}=8\cdot 10^7$ cm. $l_{min}$ and $l_{max}$ refer to the smallest and largest 
scale turbulene scale length.
We then infer for the Alfven speed $V_A=10^6$ cm s$^{-1}$, 
so that the ratio $\chi =V_A/v_e=0.05$. The non-relativistic electron and proton
gyrofrequencies are $\Omega _e=18$ Hz and $\Omega _p=0.01$ Hz, respectively.
For the thermal electron and proton gyroradii we obtain $R_e=v_e/|\Omega _e|=1.1\cdot 10^6$ cm
and $R_p=(m_p/m_e)^{1/2}R_e=43R_e=4.7\cdot 10^7$ cm.

Because one is dealing with waves representing departures from a homogenous
system, there are fluctuations in particle number density and magnetic field
fluctuations that are correlated. The basic procedure to identify these
variations has been originally given for the solar wind plasma 
by Wu \& Huba (\cite{wh75}). 
However, their analysis deals with a specific range of parameter values, in
particular they consider the case $v_e<<V_A$ opposite to ours.  
Here we calculate anew the relation between electron number density
fluctuations, $\delta n_e$, and magnetic field fluctuations, $\delta B_i$, for a more general
range of plasma parameter values.
\section{Kinetic derivation of fluctuations in the electron number density}
\smallskip
\leftline{3.1 {\it General reduction for isotropic background distribution functions}}
\smallskip
Expanding the Fourier-Laplace-transformed Vlasov
equation and Maxwell's equations for collisionless plasmas to first order in 
perturbed quantities one obtains the fluctuations, $\delta f_a$, to
the particle distribution function of plasma species $a$ and the 
magnetic field fluctuations $\delta \vec{B}$ in terms of the fluctuations 
$\vec{E}$ in the electric field for any wave type 
any isotropic equilibrium distribution function $f_a^{(0)}(p)$ 
(see e.g. Schlickeiser \cite{s02}, Eqs. (B.16) and (8.3.18))
\bdm 
\delta f_a(\vec{k},\omega ,p)=-{q_a\sa \over \Omega _a}{\gamma \over p}
{\partial f_a^{(0)}(p)\over \partial p}
\int_{-\infty \sa }^0d\alpha 
\bigl(p_{\perp }[E_x\cos (\phi -\alpha )+E_y\sin (\phi -\alpha )]
\edm
\be
+E_zp_{\parallel }\bigr)
\exp \bigl [{\imath \gamma \over \Omega _a}[(k_{\parallel }v_{\parallel }-
\omega )\alpha -k_{\perp }v_{\perp }(\sin (\phi -\alpha )-\sin \phi )]\bigr]
\label{delfa}
\ee
and
\be
\delta B(\vec{k}, \omega )={c\over \omega }\vec {k} \times \vec{E}
\label{delb}
\ee
where we have adopted a cartesian coordinate system 
with the wave vector $\vec{k}=(k_{\perp },0,k_{\parallel })=
k(\sin \theta ,0,\cos \theta )$ and the uniform background magnetic field 
$\vec{B}_0=(0,0,B_0)$ lying parallel to the $z$-axis.

By integrating $\delta f_a$ over all particle momenta 
$\vec{p}=\gamma m_a\vec{v}$, one
obtains the fluctuations in number density, $\delta n_a$, for each particle
species, thereby directly relating the number density fluctuations to the
electric field fluctuations
\bdm
\delta n_a(\vec{k},\omega )=
\int d^3p\, \delta f_a(\vec{k},\omega ,p)=
\int_0^{2\pi }d\phi \int_0^{\pi }d\Theta \sin \Theta 
\int _0^\infty dp p^2\delta f_a=
\edm
\bdm
-{q_a\sa (m_ac)^2\over \Omega _a}
\int_0^\infty dx\; \sqrt{1+x^2}x^2
{\partial f_a^{(0)}(x)\over \partial x}
\edm
\bdm
\int_{-\infty \sa }^0d\alpha 
\exp [-{\imath \sa \omega \sqrt{1+x^2}\alpha \over \Omega _a}] 
\bigl([E_x\cos \alpha -E_y\sin \alpha ]\sin \Theta \cos \phi 
\edm
\bdm
+[E_x\sin \alpha +E_y\cos \alpha ]\sin \Theta \sin \phi +E_zcos \Theta \bigr)
\edm
\be
\exp \bigl [-{\imath cx\sa \over \Omega _a}
[k_{\parallel }\alpha \cos \Theta +
k_{\perp }(1-\cos \alpha )\sin \Theta \sin \phi 
+k_{\perp }\sin \alpha \sin \Theta \cos \phi )]\bigr]
\label{deltana}
\ee
where we have introduced the normalised momentum $x=p/(m_ac)$ so that
$\gamma =\sqrt{1+x^2}$.

Applying the general integral (Gradshteyn \& Ryzhik \cite{gr65}, Eq. (4.624))
\bdm 
J(A,B,D)\equiv \int_0^{\pi} d\Theta \int_0^{2\pi }d\phi \; 
F(A\cos \Theta +B\sin \Theta \cos \phi +D\sin \Theta \sin \phi )=
\edm
\be
2\pi \int _{-1}^1dtF(Rt)=4\pi \sinh (R)/R
\label{grint}
\ee
to the function $F(t)=\exp (t)$, where $R=\sqrt{A^2+B^2+D^2}$, and derivatives of
Eq. (\ref{grint}) with respect to $A$, $B$, and $D$, respectively, we reduce 
Eq. (\ref{deltana}) to
\bdm
\delta n_a=
-{4\pi \imath cq_a(m_ac)^2\over \Omega ^2_a}
\int_0^\infty dx\; \sqrt{1+x^2}x^3
{\partial f_a^{(0)}(x)\over \partial x}
\edm
\bdm
\int_{-\infty \sa }^0d\alpha 
\exp [-{\imath \sa \omega \sqrt{1+x^2}\alpha \over \Omega _a}]
\bigl(k_{\parallel }E_z\alpha +
\edm
\be
k_{\perp }[E_x\sin \alpha +E_y(\cos \alpha -1)]\bigr)
\Bigl[{\sin Gx\over G^3x^3}-{\cos Gx\over G^2x^2}\Bigr]
\label{deltana1}
\ee
where
\be
G\equiv {c\over \Omega _a}\sqrt{k_{\parallel }^2\alpha ^2+\, 
2k_{\perp }^2(1-\cos \alpha )}
\label{Gdef}
\ee

\smallskip
\smallskip
\leftline{3.2 {\it Maxwellian background distribution functions}}
\smallskip
Adopting the thermal equilibrium distribution
\be
f_a^{(0)}={\mu _an_a\over 4\pi (m_ac)^3K_2(\mu _a)}
\exp [-\mu _a\sqrt{1+x^2}]
\label{thermaldis}
\ee
with the temperature parameter 
\be
\mu _a={m_ac^2\over k_BT_a}
\label{muadef}
\ee
Eq. (\ref{deltana1})
becomes
\bdm
\delta n_a=
{\imath  q_a\mu _a^2n_a\over \Omega ^2_am_aK_2(\mu _a)} 
\int_{-\infty \sa }^0d\alpha 
G^{-3}
\bigl(k_{\parallel }E_z\alpha +
k_{\perp }[E_x\sin \alpha +E_y(\cos \alpha -1)]\bigr)
\edm
\be 
\int_0^\infty dx\;\exp [-Q\sqrt{1+x^2}]
\Bigl[x\sin Gx-\, Gx^2\cos Gx\Bigr]
\label{deltanatherm}
\ee
where 
\be
Q\equiv \mu _a + \; {\imath \sa \omega \alpha \over \Omega _a}
\label{defQ}
\ee
The $x$-integrations can be readily performed using
(Gradshteyn \& Ryzhik \cite{gr65}, Eq. (3.914))
\be
J(G)=\int_0^\infty dx\, e^{-Q\sqrt{1+x^2}}\, \cos Gx=
Q{K_1(\sqrt{Q^2+G^2})\over \sqrt{Q^2+G^2}}
\label{intjg}
\ee
and its first and second derivative with respect to $G$ yielding
\bdm 
\int_0^\infty dx\;\exp [-Q\sqrt{1+x^2}]
\Bigl[x\sin Gx-\, Gx^2\cos Gx\Bigr]
\edm
\be
=-{\partial J\over \partial G}+\; G{\partial ^2J\over \partial G^2}
=QG^3{K_3(\sqrt{Q^2+G^2})\over (Q^2+G^2)^{3/2}}
\label{intjg1}
\ee
so that
\bdm
\delta n_a=
{\imath  q_a\mu _a^2n_a\over \Omega ^2_am_aK_2(\mu _a)} 
\int_{-\infty \sa }^0d\alpha 
\bigl(k_{\parallel }E_z\alpha +
\edm
\be
k_{\perp }[E_x\sin \alpha +E_y(\cos \alpha -1)]\bigr)
Q{K_3(\sqrt{Q^2+G^2})\over (Q^2+G^2)^{3/2}}
\label{deltanatherm1}
\ee
which is exact for all temperature values $\mu _a$.

\smallskip
\smallskip
\leftline{3.3 {\it Nonrelativistic temperature limit} $\mu _a>>1$}
\smallskip
For non-relativistic temperatures $\mu _a>>1$ and low-frequencies 
$\omega <<\Omega _a<<\mu_a\Omega _a$ we use the approximations 
$K_{\nu }(x)\simeq \sqrt{\pi /2x}\exp (-x)$ 
and 
\be
G^2<<\mu _a^2
\label{Gapp}
\ee
to derive for Eq. (\ref{deltanatherm1}) the compact expression 
\be
{\delta n_a\over n_a}\simeq 
{\imath  q_a\over \Omega ^2_am_a} 
\bigl(k_{\parallel }E_zH_1+\; 
k_{\perp }E_xH_2+\; k_{\perp }E_yH_3\bigr)
\label{nathermnon2}
\ee
It remains to solve the three 
$\alpha $-integrals 
\be
H_1=
\int_{-\infty \sa }^0d\alpha \alpha 
\exp \Bigl[-{\imath \sa \omega \alpha \over \Omega _a}-
{c^2k_{\parallel }^2\alpha ^2\over 2\mu _a\Omega _a^2}
+\rho _a(\cos \alpha -1)
\Bigr],
\label{H1}
\ee
\be
H_2=\int_{-\infty \sa }^0d\alpha \sin \alpha 
\exp \Bigl[-{\imath \sa \omega \alpha \over \Omega _a}-
{c^2k_{\parallel }^2\alpha ^2\over 2\mu _a\Omega _a^2}
+\rho _a(\cos \alpha -1)
\Bigr]
\label{H2}
\ee
and
\bdm 
H_3=\int_{-\infty \sa }^0d\alpha (\cos \alpha -1)
\exp \Bigl[-{\imath \sa \omega \alpha \over \Omega _a}-
{c^2k_{\parallel }^2\alpha ^2\over 2\mu _a\Omega _a^2}
+\rho _a(\cos \alpha -1)
\Bigr]
\edm
\be
={\partial \over \partial \rho _a}
\int_{-\infty \sa }^0d\alpha 
\exp \Bigl[-{\imath \sa \omega \alpha \over \Omega _a}-
{c^2k_{\parallel }^2\alpha ^2\over 2\mu _a\Omega _a^2}
+\rho _a(\cos \alpha -1)
\Bigr]
\label{H3}
\ee
where we have introduced the kineticity 
\be
\rho _a={c^2k_{\perp }^2\over \mu _a\Omega _a^2}={k_{\perp }^2v^2_{th,a}\over 3\Omega _a^2}
={1\over 3}k_{\perp }^2R_a^2
\label{rodef}
\ee
where $R_a=v_{th,a}/|\Omega _a|$ denotes the gyroradius of thermal particles of species $a$.

Using the identity in terms of the modified Bessel function $I_n(\rho )$ 
of the first kind,
\be
e^{\rho _a\cos \alpha }=\sum_{n=-\infty }^{\infty }I_n(\rho _a)e^{in\alpha }
\label{inident}
\ee
we obtain
\bdm 
H_1=-{\mu _a\Omega _a^2e^{-\rho _a}\over c^2k^2_{\parallel }}
\sum_{n=-\infty}^\infty I_n(\rho _a)\Bigl[1+\; 
\imath \sqrt{{\pi \mu _a\Omega _a^2\over 2c^2k^2_{\parallel }}}
\edm
\be
\exp \bigl[-{\mu _a\Omega _a^2\over 2c^2k^2_{\parallel }}({\omega \over |\Omega _a|}-n)^2\bigr]
[\sa + 
\rm{erf} \bigl(\imath \sqrt{{\mu _a\Omega _a^2\over 2c^2k^2_{\parallel }}}
({\omega \over |\Omega_a|}-n)\bigr)]\Bigr]
\label{H11}
\ee
Likewise
\bdm
H_2={e^{-\rho _a}\over 2\imath }\sum _{n=-\infty }^\infty 
I_n(\rho _a)
\int_{-\infty \sa }^0d\alpha [e^{\imath \alpha }-e^{-\imath \alpha }]
\exp \Bigl[-{c^2k_{\parallel }^2\alpha ^2\over 2\mu _a\Omega _a^2}
-\imath \alpha ({\omega \over |\Omega _a|}-n)\Bigr]
\edm
\bdm
={e^{-\rho _a}\over 2\imath }
\sqrt{{\pi \mu _a\Omega _a^2\over 2c^2k^2_{\parallel }}}
\sum _{n=-\infty}^\infty 
I_n(\rho )
\edm
\bdm
\Bigl[\sa \Bigl(
\exp \Bigl[-{\mu _a\Omega _a^2\over 2c^2k_{\parallel }^2}
({\omega \over |\Omega _a|}-n-1)^2\Bigr]-
\exp \Bigl[-{\mu _a\Omega _a^2\over 2c^2k_{\parallel }^2}
({\omega \over |\Omega _a|}-n+1)^2\Bigr]\Bigr)
\edm
\bdm
+
\exp \Bigl[-{\mu _a\Omega _a^2\over 2c^2k_{\parallel }^2}
({\omega \over |\Omega _a|}-n-1)^2\Bigr]
\rm{erf} \bigl(\imath \sqrt{{\mu _a\Omega _a^2\over 2c^2k^2_{\parallel }}}
({\omega \over |\Omega_a|}-n-1)\bigr)
\edm
\be
-
\exp \Bigl[-{\mu _a\Omega _a^2\over 2c^2k_{\parallel }^2}
({\omega \over |\Omega _a|}-n+1)^2\Bigr]
\rm{erf} \bigl(\imath \sqrt{{\mu _a\Omega _a^2\over 2c^2k^2_{\parallel }}}
({\omega \over |\Omega_a|}-n+1)\bigr)
\Bigr]
\label{H21}
\ee
Finally
\bdm
H_3=\sum _{n=-\infty }^\infty 
{\partial (e^{-\rho _a}I_n(\rho _a))\over \partial \rho _a}
\int_{-\infty \sa }^0d\alpha 
\exp \Bigl[-{c^2k_{\parallel }^2\alpha ^2\over 2\mu _a\Omega _a^2}
-\imath \alpha ({\omega \over |\Omega _a|}-n)\Bigr]
\edm
\bdm
=\sqrt{{\pi \mu _a\Omega ^2_a\over 2c^2k^2_{\parallel }}}
\sum _{n=-\infty}^\infty 
{\partial (e^{-\rho _a}I_n(\rho _a))\over \partial \rho _a}
\exp \Bigl[-{\mu _a\Omega _a^2\over 2c^2k_{\parallel }^2}
({\omega \over |\Omega _a|}-n)^2\Bigr]
\edm
\be
\Bigl[\sa +
\rm{erf} \bigl(\imath \sqrt{{\mu _a\Omega ^2_a\over 2c^2k^2_{\parallel }}}
({\omega \over |\Omega _a|}-n)\bigr)\Bigr]
\label{H31}
\ee

\smallskip
\smallskip
\leftline{3.4 {\it Electron density fluctuations}}
\smallskip
According to Sect. 2 we obtain for the electron kineticity (\ref{rodef}) 
in the diffuse intercloud medium 
\be
\rho _e={1\over 3}R_e^2k^2_{\perp }={\sin ^2\theta \over 3}(R_ek)^2\le
{1\over 3}(R_ek_{max})^2=2.6\cdot 10^{-3}<<1
\label{roelec}
\ee
values much smaller than unity at all wavenumbers and propagation angles. 
For electron fluctuations it is therefore 
justified to take the limit $\rho _e\to 0$ of  
Eqs. (\ref{H11})-(\ref{H31}).

It is convenient to introduce  
Dawson's integral (Lebedev \cite{l72}, p.19ff.)
\be
D[x]\equiv e^{-x^2}\int_0^xdt\, e^{t^2}=-{\imath \pi ^{1/2}\over 2}e^{-x^2}
\rm{erf} (\imath x)
\label{dawson}
\ee
and the parameters
\be
\psi _e^2\equiv {\mu _e\omega ^2\over 2c^2k^2_{\parallel}}=
{3\omega ^2\over 2v_e^2k^2_{\parallel}}
\label{psie}
\ee
and 
\be
\alpha _e\equiv {|\Omega _e|\psi _e\over \omega }=
\sqrt{{3\over 2}}{1\over R_ek\cos \theta }>>1
\label{alphaedef}
\ee
In the limit $\rho _e\to 0$ we obtain
\be
H_1(\rho _e=0)=\alpha _e^2\Bigl[
2\psi _eD[\psi _e]+\, \imath \pi ^{1/2}\psi _ee^{-\psi _e^2}-\; 1\Bigr]
\label{H1r0}
\ee
\bdm
H_2(\rho _e=0)={|\Omega _e|\over \omega }\psi _e\Bigl[
D[\psi _e(1-{|\Omega _e|\over \omega }]
-D[\psi _e(1+{|\Omega _e|\over \omega }]-\; 
\edm
\bdm
\imath {\pi ^{1/2}\over 2}\bigl[
\exp \bigl(-\psi _e^2(1+{|\Omega _e|\over \omega })^2\bigr)
-\exp \bigl(-\psi _e^2(1-{|\Omega _e|\over \omega })^2\bigr)\bigr]\Bigr]
\edm
\be
=-\alpha _e\Bigl[
D[\alpha _e-\psi _e]+\, D[\alpha _e+\psi _e]
-\; \imath \pi ^{1/2}\sinh ({2\alpha _e\psi _e})
\exp \bigl(-\psi _e^2-\alpha _e^2\bigr)\Bigr]
\label{H2r0}
\ee
and
\bdm
H_3(\rho _e=0)=\imath {|\Omega _e|\over \omega }\Bigl[
D[\psi _e(1+{|\Omega _e|\over \omega }]
+D[\psi _e(1-{|\Omega _e|\over \omega }]-2D[\psi _e]\Bigr]
\edm
\bdm 
+{\pi ^{1/2}\over 2}{|\Omega _e|\over \omega }\Bigl[
2e^{-\psi _e^2}-
\exp \bigl(-\psi _e^2(1+{|\Omega _e|\over \omega })^2\bigr)
-\exp \bigl(-\psi _e^2(1-{|\Omega _e|\over \omega })^2\bigr)\Bigr]
\edm
\bdm
=\imath {\alpha _e\over \psi _e}\Bigl[
D[\alpha _e+\psi _e]
-D[\alpha _e-\psi _e]-2D[\psi _e]\Bigr]
\edm
\be
+{\pi ^{1/2}\over 2}{\alpha _e\over \psi _e}e^{-\psi _e^2}
\Bigl[1-\cosh (2\alpha _e\psi _e)\exp \bigl(-\alpha _e^2\bigr)
\Bigr]
\label{H3r0}
\ee
These equations can be approximated further using the known properties of Dawson's integral.
Dawson's integral satisfies the linear differential equation
\be
{dD(x)\over dx}=1-\; 2xD(x),
\label{dawsonequation}
\ee
has a maximum $D_m=0.541$ for $x=x_M=0.924$ and an inflection point at $x=x_w=1.502$ where
$D=D_w=0.428$. At large arguments (Schlickeiser \& Mause \cite{sm95})
\be
D(x>>1)\simeq {1\over 2x}\bigl(1+\, {1\over 2x^2}+\; {3\over 4x^4}\bigr)
\label{dawsonlarge}
\ee
whereas at small arguments
\be
D(x<<1)\simeq x-\, {2\over 3}x^3+\, {4\over 15}x^5
\label{dawsonsmall}
\ee
To proceed we have to calculate the parameter $\psi _e$ from the dispersion relations of 
Alfven and magnetosonic waves in the diffuse intercloud medium.
\section{Dispersion relationships in the diffuse intercloud medium}
According to Sect. 2 the plasma parameters of the diffuse intercloud medium are 
in a range where
\be
v_p^2<<V_A^2<<v_e^2
\label{thermalvelorange}
\ee
We noted already (see Eq. (\ref{roelec}) that the electron 
kineticity is much smaller than unity.
For the proton kineticity
\be
\rho _p={m_p\over m_e}\rho _e=4.77({k\over k_{max}})^2\sin^2 \theta
\label{ropdef}
\ee
we also find values much smaller than unity if we limit our discussion to wavenumbers
much smaller than $k<<0.46k_{max}$. In this limit according 
to Sitenko (\cite{s67}, p. 115ff.) two low-frequency 
transverse plasma modes
exist: the Alfven wave and the magnetosonic wave.

The Alfven mode calculated from the dispersion relation 
$\rm {det} \Lambda _{ij}=0$ obeys 
\be
\omega ^2=V_A^2k_{\parallel }^2=V_A^2k^2\cos ^2\theta \;\;\;\;
\hbox{  Alfven mode}
\label{alfdis1}
\ee
The Maxwell operator $\Lambda _{ij} $ in the relation $\Lambda _{ij}E_j=0$ 
also specifies the wave's polarisation characteristics
\be
{E_y\over E_x}={\Lambda _{11}\Lambda _{23}+\Lambda _{12}\Lambda _{13}\over
\Lambda _{13}\Lambda _{22}-\Lambda _{12}\Lambda _{23}},\;\;\;
{E_z\over E_x}={\Lambda _{12}^2+\Lambda _{11}\Lambda _{22}\over
\Lambda _{12}\Lambda _{23}-\Lambda _{13}\Lambda _{22}}
\label{polalg}
\ee
For obliquely propagating Alfven waves Sitenko gives
\be
\vec{E}_A\simeq E_x\Bigl(1;\; -\imath {\omega \over \Omega _p\tan ^2\theta };
-{v_s^2\over V_A^2}{\omega ^2\over \Omega _p^2\sin \theta \cos \theta }\Bigr)
\label{polalf}
\ee
at frequencies much smaller than the non-relativistic 
proton gyrofrequency ($\omega << \Omega _p$).

The magnetosonic mode obeys the dispersion relation
\be
\omega ^2=V_A^2k_{\parallel }^2
+\; [V_A^2+\, v_s^2)]k_{\perp }^2
=V_A^2[k_{\parallel }^2+\; (1+\beta )k_{\perp }^2]
\;\;\;\;\hbox{  magnetosonic mode}
\label{magsodis1}
\ee
if the velocity of sound is defined by means of the equalities
\be
v_s^2={2\over 3}v_p^2\cases{
1+{T_e\over 2T_p}&   for $\cos \theta \ge \chi $ \cr
1+{T_e\over T_p} &   for $\cos \theta \le \chi $ \cr}
=\cases{
v_p^2&   for $\cos \theta \ge \chi $ \cr
{4\over 3}v_p^2        &   for $\cos \theta \le \chi $ \cr}
\label{soundspeed}
\ee
Note that the critical angle $\theta _c=\arccos (\chi )=87$ degrees.
The 
plasma beta is defined as
\be
\beta ={v_s^2\over V_A^2}={v_p^2\over V_A^2}={m_e\over m_p\chi ^2}=0.22,
\label{betadef}
\ee
The polarisation vector for obliquely propagating magnetosonic waves is
\be
\vec{E}_M\simeq E_y\Bigl(\; -\imath {\omega \over \Omega _p\sin ^2\theta }; 1;
-\imath {v_s^2\over V_A^2}{\omega \over \Omega _p}\sin \theta \cos \theta 
\Bigr)
\label{polmag}
\ee
at frequencies much smaller than the non-relativistic 
proton gyrofrequency ($\omega << \Omega _p$).
\section{Alfvenic electron density fluctuations}
Using the Alfven wave dispersion relation (\ref{alfdis1}) we obtain for Eq. (\ref{psie})
\be
\psi_{e,A}=\sqrt{1.5} \chi =0.061<<1
\label{psia}
\ee
which is much smaller than unity and much smaller than the parameter $\alpha _e$ (see Eq. 
(\ref{alphaedef}).

In the limit $\psi _{e,A}<<1<<\alpha _e$ Eqs. (\ref{H1r0})-(\ref{H3r0}) reduce to
\be
H_1^A(\rho _e=0)\simeq -\alpha _e^2,\,\,\;
H_2^A(\rho _e=0)\simeq -1,\;\;\,
H_3^A(\rho _e=0)\simeq {\pi ^{1/2}\over 2}{\alpha _e\over \psi _{e,A}}
\label{Half1}
\ee
so that according to Eq. (\ref{nathermnon2})
\be
{\delta n^A_e\over n_e}\simeq 
{\imath  e\over \Omega ^2_em_e} 
\bigl(k_{\parallel }\alpha _e^2E_z+\; 
k_{\perp }E_x-\; k_{\perp }
{\pi ^{1/2}\over 2}{\alpha _e\over \psi _{e,A}} E_y\bigr)
\label{nealfv1}
\ee
The dominant term in the bracket of 
Eq. (\ref{nealfv1}) comes from the parallel electric field ($E_z$) term
indicating that the electron density perturbation in oblique Alfven waves 
arises in response to 
their parallel electric fields.

Making use of the polarisation properties of oblique Alfven waves (\ref{polalf}) then yields
\be 
{\delta n^A_e\over n_e}\simeq 
{\imath  eE_x\over \Omega ^2_em_e}
\Bigl[k_{\perp }\; 
-k_{\parallel }\alpha _e^2{v_p^2\over V_A^2\Omega _p^2}
{\omega ^2\over \sin \theta \cos \theta }+\; 
{\imath \pi ^{1/2}\over 2}k_{\perp }{\omega \over \Omega _p\tan ^2\theta }
{\alpha _e\over \psi _{e,A}}\Bigr]
\label{nealfv2}
\ee
Using Faraday's induction law (\ref{delb}) we can express the 
magnetic field fluctuations $(B_x,B_y,B_z)$
in terms of the electric field fluctuations $(E_x,E_y,E_z)$
\be
(B_x,B_y,B_z)={c\over \omega }\vec{k}\times (E_x,E_y,E_z)
\label{faraday1}
\ee
yielding
for Alfven waves
\bdm
B_y=\pm {c\over V_A}(1+k^2R_p^2)E_x\simeq \pm {c\over V_A}E_x,\,\;\;
B_x=i{c\over V_A}
({m_p\over m_e})^{1/2}\chi kR_p{\cos \theta \over \tan ^2\theta }E_x,\;\;\;
\edm
\be
B_z=- i{c\over V_A}({m_p\over m_e})^{1/2}
\chi kR_p{\sin \theta \over \tan ^2\theta }E_x
\label{faraday2}
\ee
so that $|B_{x,z}|<<|B_y|$. Consequently,
\bdm 
{\delta n^A_e\over n_e}\simeq 
\pm {\imath  V_AkB_y\over \Omega _eB_0}
\bigl(\sin \theta -\; 
{3\over 2}{m_p\over m_e}{1\over \sin \theta }\; \pm
{\imath \pi ^{1/2}\over 2}{m_p\over m_e}{\cos ^2\theta \over \sin \theta }
\bigr)
\edm
\be
\simeq 
\mp \imath {3\over 2}{V_AkB_y\over \Omega _pB_0\sin \theta }
\bigl(1\; \mp \imath {\pi ^{1/2}\over 3}\cos ^2\theta \bigr)
\label{nealfv3}
\ee
The corresponding wavenumber power spectra are then related by 
\be
{P^A_{nn}(\vec{k})\over n_e^2}=
{9V^2_Ak^2\over 4\Omega ^2_p\sin ^2\theta }(1+{\pi \over 9}\cos ^4\theta )
{P_{yy}(\vec{k})\over B_0^2}
\label{powerrela}
\ee%
\section{Magnetosonic electron density fluctuations}
Using the magnetosonic wave dispersion relation (\ref{magsodis1}) we obtain for Eq. (\ref{psie})
\bdm
\psi _{e.M}=\sqrt{3\over 2}\chi \bigl[1+(1+\beta )\tan ^2\theta ]^{1/2}
=0.061\sqrt{1+1.22\tan ^2\theta }
\edm
\be
\simeq 
\cases{0.061 & for $\theta \le 42.2$\cr
0.068\tan \theta & for $\theta \ge 42.2$\cr}
\label{psiemv}
\ee
which is smaller than unity for propagation angles less than $\theta _c=\arctan (14.78)=86.1$ degrees.

In the limit $\psi _{e,M}<<1<<\alpha _e$ Eqs. (\ref{H1r0})-(\ref{H3r0}) reduce to
\be
H_1^M(\rho _e=0)\simeq -\alpha _e^2,\,\,\;
H_2^M(\rho _e=0)\simeq -1,\;\;\,
H_3^M(\rho _e=0)\simeq {\pi ^{1/2}\over 2}{\alpha _e\over \psi _{e,M}}
\label{Hmag1}
\ee
whereas in the limit $1<<\psi _{e,M}<<\alpha _e$
\be
H_1^M(\rho _e=0)\simeq {\alpha _e^2\over 2\psi _{e,M}^2},\,\,\;
H_2^M(\rho _e=0)\simeq -1,\;\;\,
H_3^M(\rho _e=0)\simeq -{\alpha _e\over \psi ^2_{e,M}}
\label{Hmag2}
\ee
\smallskip
\smallskip
\leftline{6.1 {\it Small and intermediate propagation angles} $\theta \le \theta _c$}
\smallskip
At propagation angles $\theta \le \theta _c$ Eq. (\ref{nathermnon2}) then becomes 
\be
{\delta n^M_e\over n_e}\simeq 
{\imath  e\over \Omega ^2_em_e} 
\bigl(k_{\parallel }\alpha _e^2E_z+\; 
k_{\perp }E_x-\; k_{\perp }
{\pi ^{1/2}\over 2}{\alpha _e\over \psi _{e,M}} E_y\bigr)
\label{nemag1}
\ee
Using the polarisation properties of oblique magnetosonic waves (\ref{polmag}) 
\be
{\delta n^M_e\over n_e}\simeq 
-{\imath eE_y\over \Omega ^2_em_e} 
\bigl(
\imath k_{\parallel }\alpha _e^2{v_p^2\over V_A^2}{\omega \over \Omega _p}\sin \theta \cos \theta 
+\; 
\imath k_{\perp }{\omega \over \Omega _p\sin ^2\theta }+\; 
{\pi ^{1/2}\over 2}k_{\perp }{\alpha _e\over \psi _{e,M}} 
\bigr)
\label{nemag2}
\ee
Faraday's induction law yields for magnetosonic waves 
\bdm
B_x=-{c\over \omega }k_{\parallel }E_y,\;\;\;\;
B_y={\imath ck\cos \theta \over \Omega _p}
[{v_p^2\over V_A^2}\sin ^2\theta -
{1\over \sin ^2\theta }]E_y,\,\;\;
\edm
\be
B_z={c\over \omega }k_{\perp }E_y
\label{faradaym1}
\ee
so that $|B_y|<<|B_{x,z}|$. 
Consequently,
\bdm
{\delta n^M_e\over n_e}\simeq 
-{\imath eB_z\over \Omega ^2_em_ec}
\bigl(
\imath \alpha _e^2{v_p^2\over V_A^2}{\omega ^2\over \Omega _p}\cos ^2\theta 
+\; \imath {\omega ^2\over \Omega _p\sin ^2\theta }+\; 
{\pi ^{1/2}\over 2}{\omega \alpha _e\over \psi _{e,M}} 
\bigr)=
\edm
\bdm
-{\imath eB_z\over \Omega ^2_em_ec} 
\bigl(
\imath {3v_p^2\cos ^2\theta \over 2R_e^2\Omega _p}[1+(1+\beta )\tan ^2\theta ]
+\; \imath {V_A^2k^2\over \Omega _p\tan ^2\theta }[1+(1+\beta )\tan ^2\theta ]\pm \; 
{\pi ^{1/2}\over 2}|\Omega _e|
\bigr)
\edm
\be
\simeq
{3\over 2}[\cos ^2\theta +(1+\beta )\sin ^2\theta ]{B_z\over B_0}=
{3\over 2}[1+\beta \sin ^2\theta ]{B_z\over B_0}
\label{nemag3}
\ee
The corresponding wavenumber power spectra are then related by 
\be
{P^M_{nn}(\vec{k})\over n_e^2}=
{P_{zz}(\vec{k})\over B_0^2}\; 
{9\over 4}[1+\beta \sin ^2\theta ]^2
\label{powerremagneto1}
\ee
As an aside we note that Eqs. (\ref{faradaym1}) imply
\be
P_{zz}(\vec{k})=\tan ^2 \theta P_{xx}(\vec{k})
\label{pzxrel}
\ee
for magnetosonic waves.

\smallskip
\smallskip
\leftline{6.2 {\it Large propagation angles} $\theta \ge \theta _c$}
\smallskip
Repeating the analysis at large propagation angles using Eqs. (\ref{Hmag2}) yields
\be
{\delta n^M_e\over n_e}\simeq 
-{B_z\over B_0}\bigl({2\over 3}({\cos \theta \over \chi })^2-\; 
(1+\beta ){m_p\over m_e}\chi ^2R_e^2k^2\bigr)
\label{nemag4}
\ee
\section{Comparison with classical MHD theory}
Eqs. (\ref{nealfv3}), (\ref{nemag3}) and (\ref{nemag4}) and consequently Eqs. 
(\ref{powerrela}) and (\ref{powerremagneto1}) are important 
modifications to the classical
MHD results (e.g. Sturrock \cite{s94}, Ch. 14.1). 

Eq. (\ref{nealfv3}) contradicts the classical MHD result 
$\delta n_e=0$ for Alfven waves. Only at wavenumbers
$k<<(2\Omega _p/3V_A)$ the kinetic result yields vanishing density flectuations
in agreement with the classical MHD theory.

For fast magnetosonic waves ($\beta <<1$) MHD theory yields
\be
{\delta n_e\over n_e}\simeq {\delta N\over N}=
{\delta B_z\over B_0}[1-{\beta \cos ^2\theta \over 1+\beta \sin ^2\theta }]^{-1}
\simeq {\delta B_z\over B_0}
\label{mhdrho}
\ee
where $N= m_pn_p+\, m_en_e$ so that $n_e\simeq  N/m_p$. Our kinetic
result (\ref{nemag3}) contains the additional factor 
\be
f_{kin}(\theta )={3\over 2}[1+\beta \sin ^2\theta ]
\label{fkin}
\ee
which increases monotonically from $f_{kin}(0)=1.5$ to $f_{kin}(90^o)=1.5(1+\beta )$.

\section{Discussion and conclusions}
Using linear kinetic plasma theory we have calculated the relation between 
electron density and magnetic field
fluctuations for low-frequency plasma waves for Maxwellian background distribution functions
of arbitrary temperatures in an uniform magnetic field. By taking the non-relativistic temperature limit 
we determined this ratio for the diffuse intercloud medium in our Galaxy.
The diffuse intercloud medium
is the dominant phase of the interstellar medium with respect to
radio wave propagation, dispersion and rotation measure studies. 
We have found differences between the relation of electron density and magnetic field
fluctuations from the linear kinetic theory as compared to the classical MHD theory.

Whereas shear Alfven waves are incompressive in MHD theory, linear kinetic theory 
yields the non-zero relation (\ref{nealfv3}) even in the limit of vanishing electron kineticity
$\rho _e=0$. Only at very small wavenumbers 
$k<<(2/3)|\Omega _p|\sin \theta /V_A=(2/3)(\omega _{p,i}/c)\sin \theta $ the kinetic result agrees
with the MHD result.

For magnetosonic waves the kinetic 
ratio of the normalised density and magnetic field fluctuations is modified from the MHD ratio 
by the factor $f_{kin}=(3/2)(1+\beta \sin ^2\theta )$ which is independent of wavenumber
and varies within values of $1.5$ and $1.5(1+\beta )$.

In the next paper of this series we
will use these kinetic turbulence relations to calculate frequency power spectra of electron density 
fluctuations 
from anisotropic power spectra of magnetic field fluctuations in the form of Alfven and
magnetosonic waves. Such anisotropic interstellar magnetic field power spectra are required in
order to be in accord with the heating/cooling balance of the diffuse intercloud medium.

\smallskip
\smallskip
{\it Acknowledgements} 
We gratefully acknowledge support by the
Deutsche For\-schungs\-ge\-meinschaft through Sonderforschungsbereich 191.

{}

\end{document}